# THE UNIVERSITY OF YORK

## Department of Computer Science

A thesis submitted in partial fulfilment for the degree of Master of Science in Advanced Computer Science

# AN EVALUATION OF PROGRAMMING LANGUAGE MODELS' PERFORMANCE ON SOFTWARE DEFECT DETECTION

Kailun Wang

Supervised by: Dr. Nicholas Matragkas

# 1 September 2019

8,968 words (excluding Reference list), as counted by Google Docs

# Abstract

This dissertation presents an evaluation of several language models on software defect datasets. A language Model (LM) "can provide word representation and probability indication of word sequences as the core component of an NLP system." [1] Language models for source code are specified for tasks in the software engineering field. While some models are directly the NLP ones, others contain structural information that is uniquely owned by source code. Software defects are defects in the source code that lead to unexpected behaviours and malfunctions at all levels. This study provides an original attempt to detect these defects at three different levels (syntactical, algorithmic and general) We also provide a tool chain that researchers can use to reproduce the experiments. We have tested the different models against different datasets, and performed an analysis over the results. Our original attempt to deploy bert, the state-of-the-art model for multitasks, leveled or outscored all other models compared.

# 1. Introduction

## 1.1 Motivation

Over 40 percent of system failures are rooted in software bugs [2]. To deliver software with higher quality and lower cost is a vital and long-lasting task. However, major tools to detect bugs like Static Bug Finders (SBFs) use traditional statistical methods, produce unsatisfying precision/recall rate, and are not fully automatic. Recent years have seen a rapid evolution of language models in the natural language processing (NLP) domain. As these models make their appearance in the natural language field, they are also deployed in the software engineering context for different tasks, since many tasks across natural language and programming language domain are similar: context analysis, classification, summarization, translation, deduction etc.

Research has already taken place to fill the gap to improve bug detection mechanism to match the level of current NLP studies [15]. As these existing models vary in structure and particular task they apply to, it became our interest to evaluate and compare their performance in the bug detection task. We also noticed that software defection is a rather big topic with many different contexts. We therefore also provided an attempt to classify these defect contexts to study them separately. In this research, we will select representative models that have proved to be fit for classification problems. We then provided an original taxonomy for software defects into three levels and set a dataset for each of these levels. We then re-implement the existing models on these three datasets to evaluate their performance. In this research, we will select some of the representing state-of-the-art models from both NLP achievements and structures in the SE field. We will then evaluate existing models over datasets of software defects. Then we will prepare our own version of datasets in case it's needed, and modify and evaluate existing language models to fit the dataset. Our core task would be inspecting whether any general information and shared properties exist between the different datasets/problems. Then we will give an evaluation to the overall result. We also provide our own attempt to solve all three tasks with one single state-of-the-art model. Chapter 5 lists a set of more specific Research Questions and Chapter 8 provides our findings to these questions.

## 1.2 Report structure

The structure of this dissertation is as follows. Chapter 2 gives an overview literature review over existing studies of the research topic: covering NLP, language structures in the SE field and our attempt to classify software defects. Chapter 5 lists our goals for this research. Chapter 6 briefly introduces our toolchain. Chapter 7 records our experiment process in detail. Chapter 8 summons up the results and checked whether each goal is met.

# 2. Research Literature

## 2.1 Natural Language Processing

Natural Language Processing contains a series of tasks relating to human language, including speech recognition, natural language understanding, and natural language generation.

The earliest research work in natural language understanding was machine translation [3]. In 1949, American Weaver first proposed a machine translation design. In the 1960s, people had a large-scale research work on machine translation, which cost a lot of money. However, people obviously underestimated the complexity of natural language. The theory and technology of language processing were not hot, so the progress was not great. The main method is to store the two-language words and phrases corresponding to the translation of the large dictionary, one-to-one correspondence in translation, technically just adjust the order of the language. But the translation of language in daily life is far from being so simple. Many times, we have to refer to the meaning of a sentence before and after. Since the beginning of the 1990s, great changes have taken place in the field of natural language processing. Two distinct features of this change are:

(1) For system input, the natural language processing system required to be developed can handle large-scale real texts, rather than dealing with very few terms and typical sentences, as in previous research systems. Only in this way can the developed system have real practical value.

(2) The output of the system is very difficult in view of the true understanding of natural language. It is not required to have a deep understanding of natural language text, but it is necessary to extract useful information from it. For example, automatic extraction of index words, filtering, retrieval, automatic extraction of important information, automatic summarization, etc., for natural language text.

At the same time, due to the emphasis on "large scale" and the emphasis on "real texts", the basic work of the following two aspects has also been emphasized and strengthened.

(1) Development of large-scale real corpus. The large-scale corpus of real texts processed through different depths is the basis for studying the statistical properties of natural language.

(2) The preparation of large-scale, informative corpuses. The scale of tens of thousands, hundreds of thousands, or even hundreds of thousands of words, the richness of information (such as the collocation information of words) is very important to the natural language processing.

## 2.2 Bug detection

**Static Bug Finders**. Existing tools like PMD [4] and Findbugs [5] detect fix patterns, algorithms as well as semantic and syntactic properties. Figure 1 shows a report generated by findbugs, in which many of these bug categories like "Unused field" and "Confusing method names" can be judged by rather fixed rules. Rahman et al. [6] summarizes the role of SBFs. The advantage is that they "scale well" and are widely adopted in the industry. The disadvantage of static bug-finding is that in a real production case, they generate many false positives as well as false negatives, which results in manual check time or an undetected bug. Also, upon the warnings given by SBFs, a developer have to check manually.

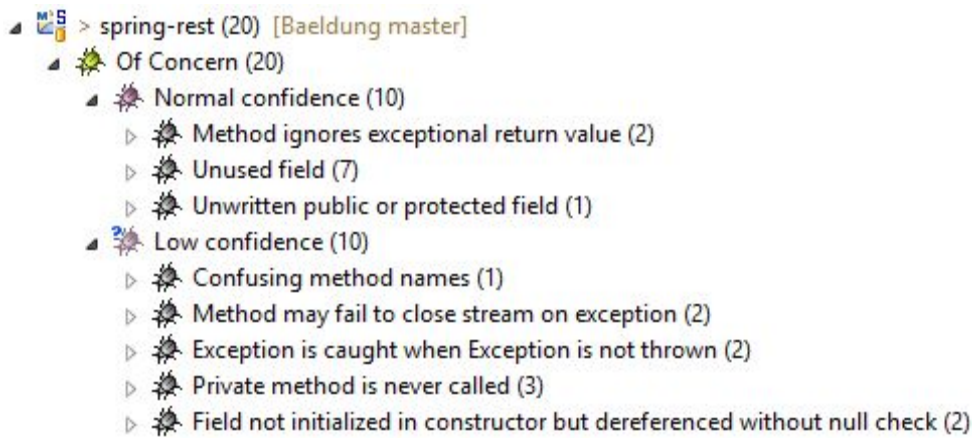

Fig 1. A report generated by FindBugs

## 2.3 Stochastic changes in bug fixes

Previous researches are also inspired by the fact that program and human language are similar distribution sequences. In terms of bug detection, the findings suggest that bug code are "less natural" judged by trained models, and that bug detection using this metric achieves similar accuracy as traditional Static Bug Finders (SBFs) . Study by Ray et al. [7] over repository commits suggests that programming languages share many common properties with human natural language, and that there's a mathematical basis on why language model would work over bug detection.

**Source code "Naturalness".** By measuring the perplexity over n-gram models, the calculated result suggests that comparing to human language, source code has less complexity, thus requires less complicated structure to captures its features.

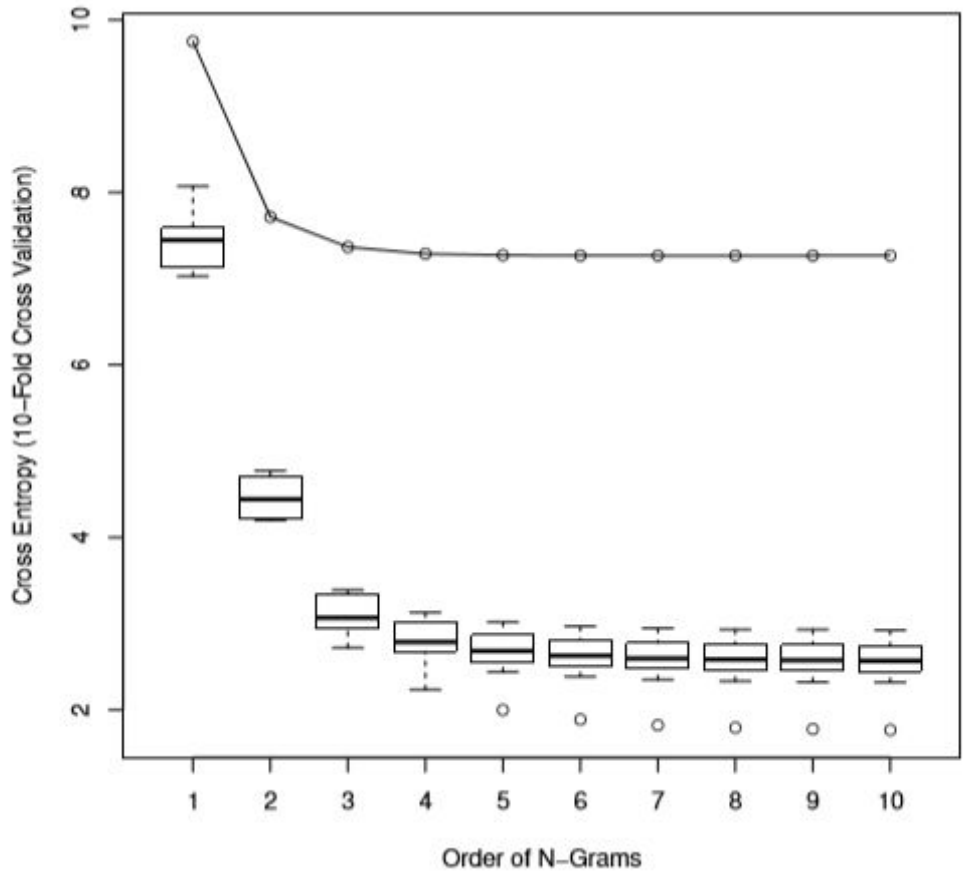

Fig 2. Cross-entropy of English and Java Repositories [8]

**Code repository "Localness".** In a certain repository, the writers inspected how repeatedly the code is distributed. They concluded that "Source code is locally repetitive, viz. it has useful local regularities that can be captured in a locally estimated cache and leveraged for software engineering tasks."

**Bug fix "Naturalness".** It is concluded in the paper that "Buggy code is rated as significantly more "unnatural" (improbable) by language models. This unnaturalness drops significantly when buggy code is replaced by fix code. "
Such findings suggest that it is mathematically possible to locate bugs through probability models, and it is also possible to catch bugs in large repositories by only inspecting a small number of local files.

## 2.4 Language Models for programming languages

### 2.4.1 Language Models

Deploying language models for software engineering tasks has almost the earliest attempts as deploying these models to natural language tasks. The models started from probability models (e.g. Tf-idf [9], to the later n-grams[10] and then common models used in NLP tasks (CNN, LSTM etc.). Recently there are attempts to study these tasks in a general model by increasing the size of the dataset to relatively large (Big code studies, fine-tuning pretrained language models for specific tasks). [11] and [12] provided a great summary on these existing models' applications in the software engineering field.

### 2.4.2 Probability models

**Bag of N-grams model.** N-gram refers to the combination of token sequences. For example, in the sentence "How old are you?", "How old are" and "old are you" are the two 3-grams. The "bag of N-grams" refers to the conditional probability of N-gram sequences, in the example being Prob("How old are you") = Prob("old are you" | "How old are") * Prob("old are you"). This is still a widely adapted method, and is used in calculating code perplexity introduced in later chapters. [10]

**Topic Model.** Latent Dirichlet Allocation (LDA) is a typical topic model which assumes an article is a probability distribution of hidden topics, and each hidden topic is a probability distribution of words. Given the set of articles with words, it is possible to calculate back the parameters of the two probability models. [13]

### 2.4.3 Linear models

**TextCNN model.** Presented by Kim. [14], this model was the state-of-the-art method for text classifications for a long time since its debut. As shown in figure 3, it uses a series of filters of different lengths to capture the different level of relation in the sentence. In the study by [15], this model has been proven to be suitable for detecting bug commits. In this experiment, we will implement a Bi-CNN structure with each CNN corresponding to one side of the input data.

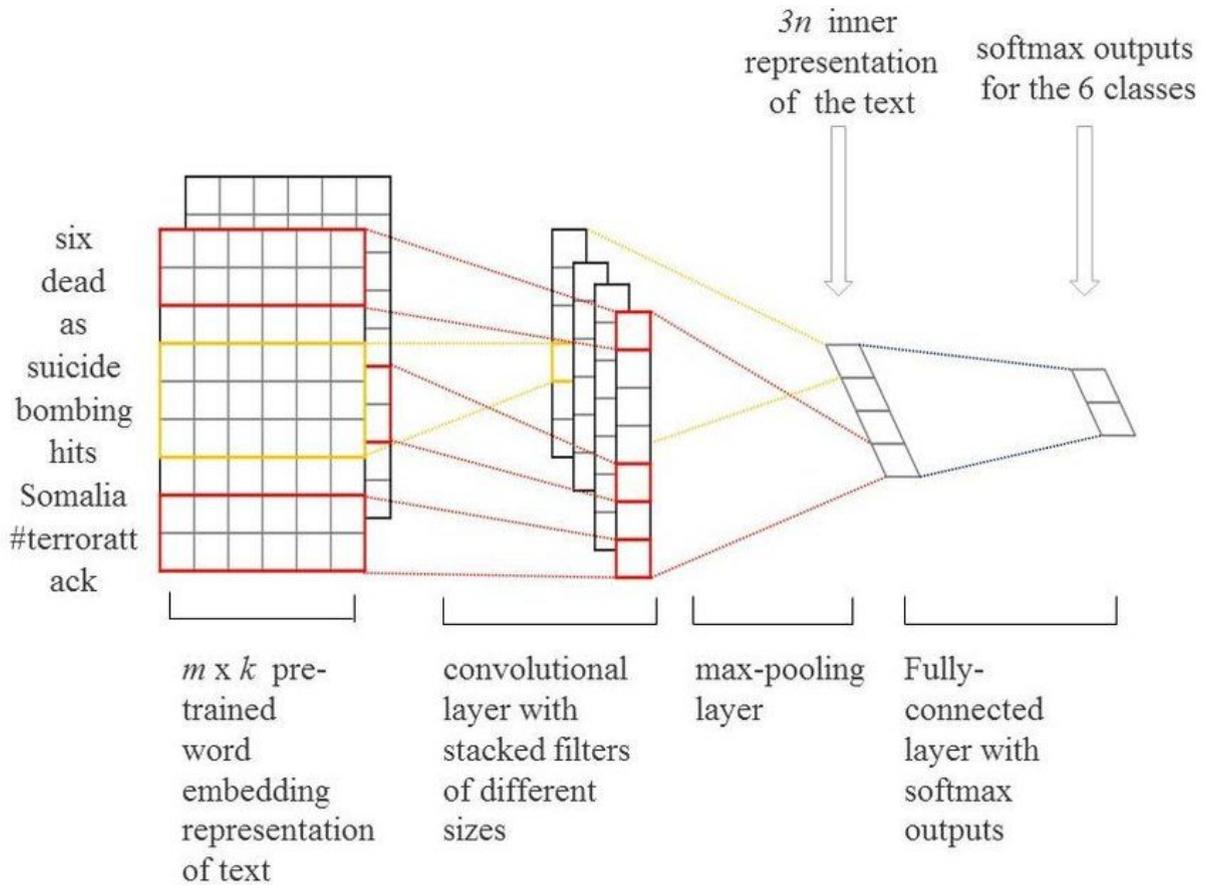

Fig 3. Original concept of Text CNN filters by Kim. [16]

**LSTM & attention model.** LSTM structure [17] is a enhancement of previous RNN structures to prevent the weights of earlier nodes being ignored. As is shown in figure 4, for a sequence input of fixed length, each unit in the sequence is corresponded with a LSTM unit. By several inside components, the unit is guided to "remember" or "forget" information from previous information and sends this "current memory" to the subsequent layer. In a study by [18], the authors deployed the LSTM model to recognize token errors that lead to compilation errors.

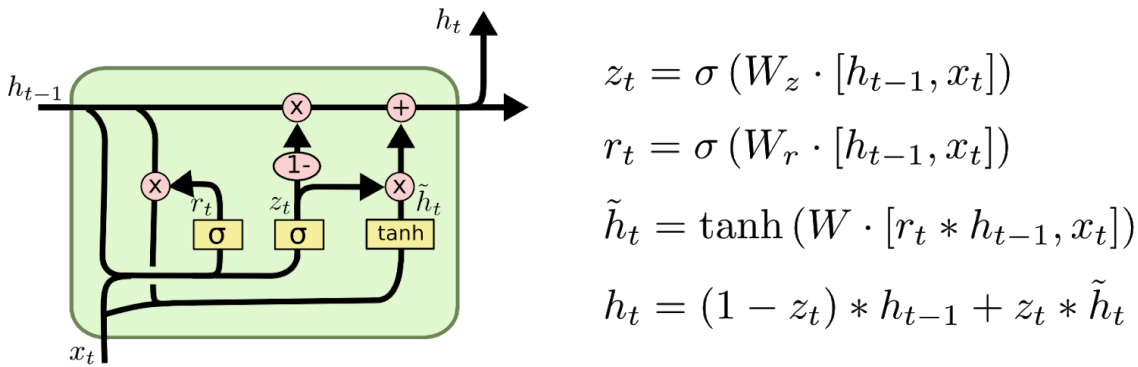

$$z_t = \sigma\left(W_z \cdot [h_{t-1}, x_t]\right)$$
$$r_t = \sigma\left(W_r \cdot [h_{t-1}, x_t]\right)$$
$$\tilde{h}_t = \tanh\left(W \cdot [r_t * h_{t-1}, x_t]\right)$$
$$h_t = (1 - z_t) * h_{t-1} + z_t * \tilde{h}_t$$

Fig 4. Basic unit in LSTM layer. [19]

The attention concept was proposed and refined by [20] and [21] to solve sequence to sequence problems. It has also inspired the famous "transformer" structure which is popular in recent studies.

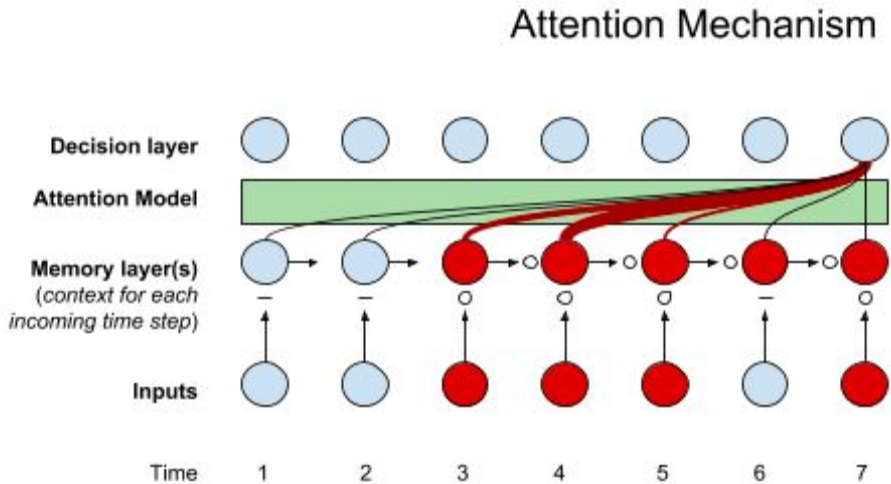

Fig 5. Attention Mechanism. [22]

In this experiment a baseline LSTM+Attention model is implemented. For each input in the input pair there's a corresponding LSTM+Attention structure.

### 2.4.4 Tree/graph models

One of the characteristics of natural language is that the conversation context proceeds in a chronological order. However, in a source code file, the context of how part of the code is related to another is more complicated. To address this issue, we also reviewed and examined structures that represent this information.

**Tree-based CNN model (TBCNN)** Presented by [23], this model further develops the original concept of the TreeCNN model by [24], which is also a study over source code. The model aims to solve the algorithm classification task across different languages (C++ and Java files implementing different algorithms like BFS, DFS, bubble sort etc.). As seen in figure 6, to abstract the entire code, the model aggregates the node level embedding vectors along the abstract syntax tree (AST) of the source code. The tree of aggregated vectors is then pooled into one single vector. In this experiment a bi_tbcnn structure is implemented for the input pairs.

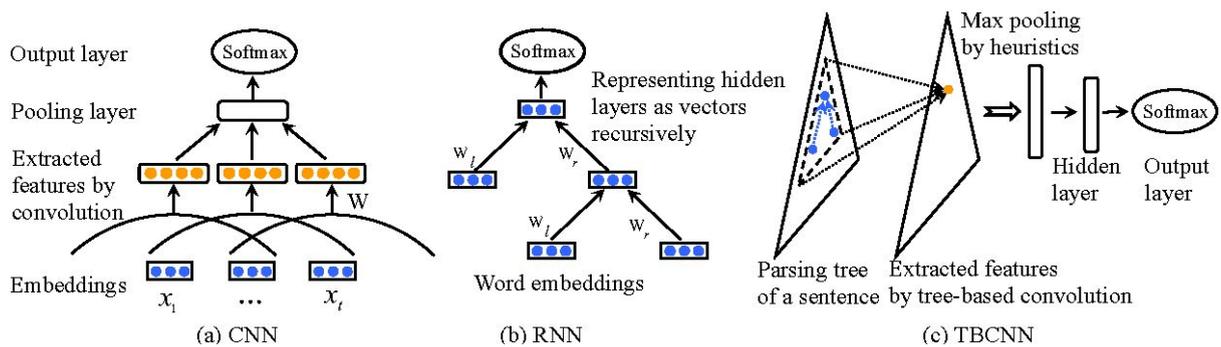

Fig 6. Traditional CNN, RNN and Tree-based CNN (TBCNN) structures compared. source [Mou, L., Peng, H., Li, G., Xu, Y., Zhang, L., & Jin, Z. (2015). Discriminative Neural Sentence Modeling by Tree-Based Convolution. EMNLP.]

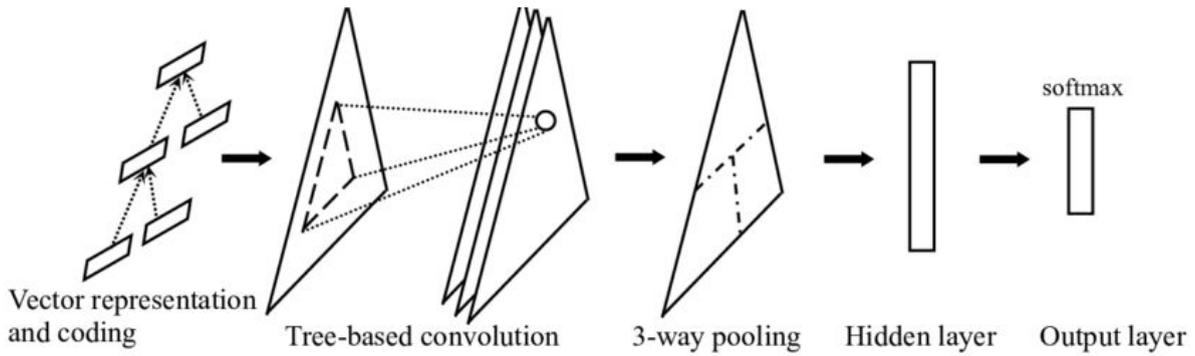

Fig. 7 General structure of a TBCNN model.[24]

**Gated Graph Neural Network (GGNN).** Presented by [25], this model is based on the original study of GGNN by Li et al. [26]. The model aims to solve the "VarMisUse" task, in which some variables are mistyped as tokens of the same class (e.g. wrong Boolean variable *hasGlobalEvent* for correct one *hasLocalEvent*). To conclude the model, it propagates the calculated weights from the upper node to the later one along the edges of a directed acyclic graph (DAG). In this problem, the DAG is a collection of edges representing the definition, usage and modification of certain vectors, but in theory, the DAG can be of any size. The propagation, unlike the ones implemented in network flow algorithms, terminates after a limited number of steps. Given that there might be special alignment issues and memory limitation, a similar bi-ggnn structure is not deployed. Fig 8 shows an application of GGNN layer following subsequent layers, but in this experiment we would just use the first GGNN layer.

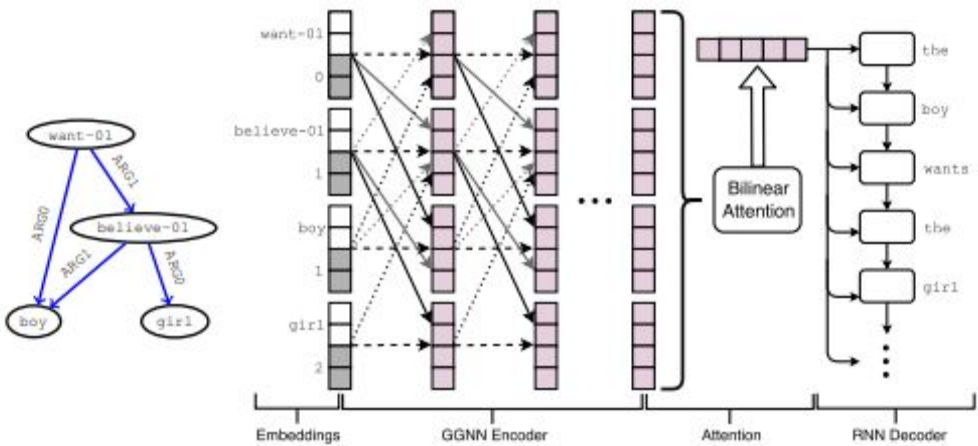

Fig 8. An application of GGNN Layer. [27]

### 2.4.5 General Language Models

The concept of training a universal model for different tasks is first brought out by OpenAI's GPT model [28]. This was then extended to the famous "BERT" model by Google[29]. Recent variations include Facebook's Roberta [30], openai's GPT-2 [31] etc. Being the state-of-the-art approach on NLP, these models have topped the leaderboard of various NLP tasks.

Deploying a general LM involves two stages: pre-training and fine-tuning. Pre-training is often done by masking some parts of the sentence and ask the model to perform a completion. Some pre-training tasks, such as the one in bert, also involves predicting the next sentence. It is worth noticing that the corpus used by the pre-training stage is relatively large. Fine-tuning applies the same model to a transfer learning task with on-demand IO format (single sequence or double sequence as input, and single label/multi label/sequence as output).

In this research we managed to apply facebook's torch implementation of Bert[1] over the three tasks.

## 2.5 Levels of source code defects

While the studies of software defects have been taken separately, this research provides a first attempt to evaluate these defects of different complexity levels into a whole. Although it is not carried out in this experiment (these cases are trained separately), we are suggesting that maybe a transfer-learning process might be applied from lower-level problems to higher-level ones. In this chapter we will introduce our taxonomy, and their our selected datasets corresponding to these levels will be described in the next chapter.

### 2.5.1 Syntactical defects

This is the most basic level of programming faults, and are easily committed by novice programmers. Such programs fail the compilation process. Examples of this level include: mismatching brackets, wrong if/for statements, undefined identifiers, mismatched classnames/numerical types etc.

**Case Example.** The following Java code has an undefined variable k and thus fails the compilation.

```
public class MyClass {
   public static void main(String args[]) {
     int x=10;
     int y=25;
     int z=x+y;
```

---

[1] https://github.com/facebookresearch/XLM

```
    System.out.println("Sum of x+y = " + k);
  }
}
```

### 2.5.2 Algorithm level

In programming contests, contestants are asked to submit a program to solve a specific algorithm problem turning given input data to designated output data. An algorithm level fault program would pass the compilation but fail some of the test cases (generated wrong output or exceeded time/memory limits).
Examples of this level of defects include: wrong boundaries, wrong operators (> for >=), undealt scenarios etc.

**Case Example.** The following code[2] from the codeforces site is a "Wrong Answer" submission for an algorithm problem. The site has suggested that an "out of boundary error" in the highlighted line led to the mistake.

```
#include <bits/stdc++.h>
using namespace std ;
int n;
double x , y , theta, max_angle ,PI = acos(-1) ;
vector < double > angle ;
void go (double l , double r) {
 if (l >= 0 && r >= 0) {}
 else if (l >= 0 && r <= 0) {theta = 360 - theta ;}
 else if (l <= 0 && y >= 0) {theta = 180 - theta ;}
 else {theta = 180 + theta ;}
 angle.push_back(theta) ;
}
int main(){
  scanf("%d" , &n) ;
  for (int i = 0 ; i < n ; i++) {
    scanf("%lf%lf" , &x , &y) ;
    if (x == 0) {
       if (y > 0) theta = 90 ;
       else theta = 270 ;
    }
    else if (y == 0) {
       if (x > 0) theta = 0 ;
       else theta = 180 ;
    }
```

---
[2] https://codeforces.com/contest/257/submission/58881110

```
  else theta = atan( abs(y) / abs(x) ) * 180.000 / PI ;
   go(x , y) ;
 }
 sort (angle.begin() , angle.end()) ;
 angle[n] = angle[0] + 360 ;
 for (int i = 1 ; i <= n ; i++)
 max_angle = max (max_angle , angle[i] - angle[i - 1]) ;
   printf("%.9lf\n" , 360 - max_angle) ;
       return 0 ;
}
```

### 2.5.3 General Bug Fixes in Commits

Among git commits of software repositories, there are ones related to bug fixes, and other ones not (e.g. adding a feature). In this research this is the most complicated level of defects. For this defect level, the context it is relevant to the largest (the modified source files are large, and the sizes of the related source code files are also the largest).

**Case example.** The following code is a part of a Java commit that fixes a certain issue in the facebook android sdk package. Both versions pass the compilation and achieves most required features, but the old one would generate a runtime error. It is also worth noting that these modifications may occur in several files.

```
v 12 ▪▪▪▪  facebook-core/src/main/java/com/facebook/FacebookSdk.java

         @@ -630,11 +630,13 @@ static void publishInstallAndWaitForResponse(
630  630            // send install event only if have not sent before
631  631            GraphResponse publishResponse = publishRequest.executeAndWait();
632  632
633       -         // denote success since no error threw from the post.
634       -         SharedPreferences.Editor editor = preferences.edit();
635       -         lastPing = System.currentTimeMillis();
636       -         editor.putLong(pingKey, lastPing);
637       -         editor.apply();
     633  +         if (publishResponse.getError() == null) {
     634  +            // denote success since there is no error in response of the post.
     635  +            SharedPreferences.Editor editor = preferences.edit();
     636  +            lastPing = System.currentTimeMillis();
     637  +            editor.putLong(pingKey, lastPing);
     638  +            editor.apply();
     639  +         }
638  640         } catch (Exception e) {
639  641            } catch (Exception e) {
640  642               // if there was an error, fall through to the failure case.
```

Fig 9. A bug fix commit in real world.

## 2.6 Datasets

### 2.6.1 General defects dataset

The paper by [7] studied common properties in repository commits. The result shows that there is a decrease of perplexity by the measure of n-gram probabilities. However it is not available for us to use this property to detect whether a fix is bug or not because the decrease of perplexity is an aggregate property when these commits are inspected as a group.

As there was not an existing copy of the dataset, we have recreated the dataset from the methodology described by RAY et al [7]. As implemented in this study, 10 java repositories with over 1,000 commits are selected (mostly are from the apache foundation and android related libraries). The method to determine bug commits [32] is then deployed to these repositories' commits to generate the bug and non-bug commits. In short, this method judges whether the keyword set including words like "defect", "bug" etc. are in the commit's description sentence. In the following experiments, we generated 2,000 pairs of data for bug and non-bug commit pairs. We also set a limit on these pairs that they should not be too long or too short, and the modification should also not be too large. These pairs, labeled with number 0 and 1, are then randomly shuffled and splitted into train/validation/test pairs (by the ratio of 8:2).

| Ecosystem | Project | Study Period | Snapshots | #Files | NCSL | # of Changes | # of Bugs |
|---|---|---|---|---|---|---|---|
| Github | Atmosphere | May-10 to Jan-14 | 17 | 17,206 | 6,329,400 | 2,481 | 1,130 |
| | Elasticsearch | Feb-10 to Jan-14 | 17 | 103,727 | 22,156,904 | 4,922 | 1,077 |
| | Facebook-android-sdk (fdk) | May-10 to Dec-13 | 16 | 3,981 | 1,431,787 | 320 | 143 |
| | Netty | Aug-08 to Jan-14 | 24 | 57,922 | 12,969,858 | 3,906 | 1,485 |
| | Presto | Aug-12 to Jan-14 | 7 | 23,086 | 6,496,149 | 1,635 | 330 |
| Apache | Derby | Sep-04 to Jul-14 | 41 | 143,906 | 61,192,709 | 5,275 | 1,453 |
| | Lucene | Sep-01 to Mar-10 | 36 | 47,270 | 11,744,856 | 2,563 | 469 |
| | OpenJPA | May-06 to Jun-14 | 34 | 131,441 | 27,709,778 | 2,956 | 558 |
| | Qpid | Sep-06 to Jun-14 | 33 | 94,790 | 24,031,170 | 3,362 | 657 |
| | Wicket | Sep-04 to Jun-14 | 41 | 159,332 | 28,544,601 | 10,583 | 994 |
| Overall | | Sep-01 to Jul-14 | 266 | 782,661 | 202,607,212 | 38,003 | 8,296 |

Table 1. The 10 source repositories that forms the dataset. source [ray et al. 2016]

### 2.6.2 The codeflaws dataset

Presented by [33], this dataset collects programmer submissions on the codeforces, a site that holds competitive programming contests and practice problems. To solve a specific problem, contestants will make several submissions to pass all the test cases (each with different inputs). The datasets contains such submission pairs, in which one of the files is a failed attempt (passes some of the test cases but not all), and the other one being the accepted (a.k.a. "AC") attempt (passes all the test cases without exceeding the designated time and memory limit). These pairs are then classified into 21 types of faults (wrong boundary, wrong loop etc.). The authors also included a set of automated code fix tools to evaluate their code fix performance.

TABLE II: Our Defect Class Classification and Examples of Each Defect Class

| AST Type | Defect Type | Defect Class | Example |
|---|---|---|---|
| Statement | Control flow | (SDIF) Delete if, else, else if, for or while | − if (lines[i].y1 == last->y1) |
| | | (SIIF) Insert if, else, else if, for or while | + if(*l*) |
| | | (SRIF) Replace if, else, else if, for or while | − if(a==b)<br>+ if(mask(a)==b) |
| | | (SDRT) Delete return | − return 0; |
| | | (SIRT) Insert return | + return 0; |
| | | (SDIB) Delete/Insert break or continue | − break; |
| | Data flow | (SDLA) Delete assignment | − answer+=((i-1)*dif); |
| | | (SISA) Insert assignment | + t=0; |
| | Function call | (SDFN) Delete function call | − printf("%s %s\n",s1,s2); |
| | | (SISF) Insert function call | + scanf("%d", &n); |
| | Type | (STYP) Replace variable declaration type | − int a;<br>+ long a; |
| | Move | (SMOV) Move statement | − scanf("%d",&i);<br>  scanf("%s", &a);<br>+ scanf("%d",&i); |
| | | (SMVB) Move brace up/down | − }<br>  printf("%d",c);<br>+ } |
| Operator | Control flow | (ORRN) Replace relational operator | − if(sum>n)<br>+ if(sum>=n) |
| | | (OLLN) Replace logical operator | − if((s[i] == '4') && (s[i] == '7'))<br>+ if((s[i] == '4') \|\| (s[i] == '7')) |
| | | (OILN) Insert && (tighten condition) or \|\| (loosen condition) | − if(t%2==0)<br>+ if(t%2==0 && t!=2) |
| | | (OEDE) Replace = with == or vice versa | − else if(n=1 && k==1)<br>+ else if(n==1 && k==1) |
| | | (OICD) Insert a conditional operator | − printf ( "%d\n", i );<br>+ printf ( "%d\n", 3 == x ? 5 : i ); |

Table 2. Partial table of defect class classifications from the codeflaws dataset. [43]

### 2.6.3 The AtCoder Beginner's Problems dataset

The blackbox dataset [34] is scheduled to be selected in the first place, but with limited time, the attempt to reconstruct the novice errors from the database is not completed. We have instead constructed a similar database over novice programmers' mistakes from an online programming contest site AtCoder (https://atcoder.jp/). We retrieved over 2,000 Java source files that failed the compilation process, and equally 2,000 files that passed the tests. We then select the line that leads to the compilation error from the faulty files and a random line whose length is over 7 tokens from the clean files, accordingly. These one-lines and the complete files form the input pairs. The scraping code is available in our toolchain, and it is easy to construct a similar database for other languages like Python, C++ etc..

# 3. Objectives

The research aims to evaluate existing models' performance on datasets of different levels of software defects, namely three tasks: syntactical defect, algorithm malfunction defect and general bug commit in git commits. We will measure a model's performance by terms of test accuracy after the train-validation period is complete. Given the results, an analysis over the properties of each dataset, the properties of problem types each of the datasets is representing and whether each model's structure matches the problem at an acceptable level.

More specifically, we have set the following **research questions** for our experiment:

> RQ1. Are generally adapted language models in natural language tasks fitted for programming languages (in terms of detecting defects)?

This question targets the "linear models": Text CNN and LSTM. Accordingly, we ask a similar question to the tree and graph models:

> RQ2. Are models designated for algorithm classification tasks fitted for defect classification task?

As we are creating some of our datasets from scratch, we would also examine whether they have reached a satisfying level for academic reference:

> RQ3. For originally generated datasets, are they general enough to represent the overall feature of the task?

Our main goal would be investigating whether different levels of defects can be treated as a similar task.

> RQ4. Is there any common ground and similarity between the different level of defects, so that a certain model is suited for all three tasks?

Finally, in terms of reproducing our research as well as deploying it for educational, research or business purposes, the performance and efficiency of the models are also vital:

> RQ5. What is the performance of the training process in each task measuring by time and space efficiency?

# 4. Toolchain & Manual

We have developed the dataset preprocessing tool specialized for each model/dataset pair. We have also modified the code in each model to fit with different datasets (binary/multi classification variation). The code repository is available on Github[3]. More implementation details on the per-problem and subtask level are described in the next section.

The repository contains several markdown files, each listing the precise commands on how to reproduce the training process, as well as the set of modified files in the original models. Our generated AtCoder dataset is hosted in a separate repository for public review ([https://github.com/hiroto-takatoshi/atcoder_java](https://github.com/hiroto-takatoshi/atcoder_java)).
Our bert repository with download links is forked from facebook's original XLM code, and it's available at (https://github.com/hiroto-takatoshi/XLM).

---

[3] https://github.com/hiroto-takatoshi/ProgLMBug

# 5. Experiements

## 5.1 Environment

All experiments are carried out on the author's personal laptop, with 16GB RAM, core i7 7th generation processor and GTX 1050Ti graphic card. As part of the original TreeCNN implementation, the GNU Parallel tool [35] is used. Part of the TreeCNN model involves using the docker image of containing the FAST parsing tool [36]. Experiment records suggested that at least 8 GB memory should be allocated to the docker VM.

## 5.2 The general commit defects dataset

### 5.2.1 TextCNN & Attention models

The original model would take a pair of token sequence as input and a label as the output. As described in the following subsections, the original code is modified to fit the requirements.

**Filtering out the bug commits.** Python's difflib library is used to generate the diff pairs between two source code files. We also modified the dictionary building process to have a compare of whether using specific token types or use the "unknown" token for less frequent tokens have a better result.

```
from git import *
import os
import difflib
import re

the_dir = "C:/Users/admin/testfolder/netty"

repo = Repo(the_dir)
assert not repo.bare
assert os.path.isdir(the_dir)
assert os.path.exists(the_dir)

def is_bug(msg):
    mst = msg.lower()
    stem_words = ['error','bug','fix','issue','mistake','incorrect','fault','defect','flaw','type']
    for _ in stem_words:
        if _ in msg: return True
```

```
   return False

bugcommitlist = []
for c in l:
   if is_bug(c.message):
      bugcommitlist.append(c)

# pick a certain bug commit
commit = bugcommitlist[444]
parent_commit = bugcommitlist[444].parents[0]
the_diff_list = commit.diff(parent_commit)

for diff_item in the_diff_list.iter_change_type('M'):

   fileA = diff_item.a_blob.data_stream.read().decode('utf-8')
   fileB = diff_item.b_blob.data_stream.read().decode('utf-8')

   diff = difflib.unified_diff(fileA, fileB)
   for line in list(diff):
      if line.startswith("@@"):
         a,b,c,d= map(int, re.findall(r'\d+', line))
         a,b,c,d = a,a+b-1,c,c+d-1
         print(a,b,c,d)
```

where a,b,c,d are the specific "changed" lines' starting number and ending numbers that would appear in a common diff tool.

**Generating the embedding vector.** We use the Github Java corpus [37], which is also used in previous studies to generate similar vectors. This dataset covers a large range of repos on GitHub that varies in size, from small tools to large platform structures, and is thus representative of the overall feature of Industrial level Java source code.

|  | Train | Test |
|---|---|---|
| Number of Projects | 10,968 | 3,817 |
| LOC | 264,225,189 | 88,087,507 |
| Tokens | 1,116,195,158 | 385,419,678 |

Table 3. Stats of the Github Java Corpus. [38]

One of the challenges is the limited number of vocabulary for the word2vec embeddings. In the original code and common practice, only the first 100,000 most frequent tokens are selected in the dictionary. For the rest of the less frequent tokens, given that in a million level corpus, the number of such tokens is huge as well, two options are made possible in the test. One is to label these tokens as a single label "<unknown>", the other option is to replace these tokens with their token types (e.g. Identifier, class name, function name etc.).

```python
def filter_token(token):
    numlist = [javalang.tokenizer.DecimalFloatingPoint, javalang.tokenizer.BinaryInteger, \
            javalang.tokenizer.DecimalInteger, javalang.tokenizer.FloatingPoint,\
            javalang.tokenizer.HexFloatingPoint, javalang.tokenizer.HexInteger,\
            javalang.tokenizer.Integer, javalang.tokenizer.OctalInteger]
    if type(token) in numlist: return "<num>", False
    elif type(token) == javalang.tokenizer.String: return "<str>", False
    else: return token.value, True

def tokenize(file):
    global vocabulary
    lines = file.read()
    try:
        tokens = javalang.tokenizer.tokenize(lines)
        for token in tokens:
            res, storeType = filter_token(token)
            vocabulary.append(res)
            if storeType: vocabulary.append(str(type(token)))
    except javalang.tokenizer.LexerError:
        print("Could not process " + file.name + "\n" + "Most likely are not Java")
```

The code above shows the modified tokenizer to replace strings and numericals.

**Input and output format:** without special explanation, all of the models below would take pairs of tokenized code as input, and generate a corresponding classification label (binary 0 or 1). The model in the middle can be viewed as a black box.
Both the TextCNN and the LSTM model achieved same 73% test accuracy on same repositories, and 48% accuracy on commits from repository different from the training ones. This suggested that either the dataset of 10 repositories is not general enough to represent the common aspects of bug commits, or that the original TextCNN model does not capture enough characteristics among these bug commits. We then changed the representation of less frequent tokens from "<unknown>" to their following types, and this resulted in a slightly drop of final test accuracy (about 2% in terms of absolute accuracy). This suggested that in contrast, detailing these tokens' information does not lead to better accuracy, but leaves a higher complexity for the model to judge. In the later experiments, the "<unknown>" label will be the default representation of the less frequent tokens.

The implementation of the baseline LSTM+Attention model is a few lines different from the TextCNN model simply by replacing the corresponding Keras layers, and is in the same file (A python notebook file) as the TextCNN implementation.

### 5.2.2 TreeCNN model

The original TreeCNN code is developed in Python 2. We used Python's built-in 2to3 and autopep8 tool to change the code from Python 2 environment to Python 3. As the original code heavily relies on the parsing tool written by the same author, there are some parts of the code that has to be run inside the docker image provided by the authors, and we have explained in the sample which line is to be run in docker and which ones are not. We have also modified the code of preprocessing the dataset to the format the model recognizes. Regarding the issue of source files being too large, it seems that it is okay to feed only part of the complete source code file into the parser without generating any errors, and we follows this practice to feed only the hard parts. Finally, we added what is called the "ReduceonPlateau" callback feature [39] in the Keras platform to the tensorflow based code to reduce the learning rate once the validation accuracy does not increase.

```
if epoch > 0 and epoch % 30 == 0:
    valid_acc = run_validation()
    if valid_acc <= previous_loss:
        rounds_no_improvement += 1
        if rounds_no_improvement > 3:
            print("no further improvements, ending...")
            break
        lr = lr * 0.5
        print("learning rate reduced to ", lr)
    else:
        rounds_no_improvement = 0
        previous_loss = valid_acc
```

where run_validation() is a function similar to the original test function that returns an in-time accuracy of the current epoch model. When the accuracy of the current epoch is no longer improving, the learning rate will be reduced by half.

The tree model achieved a 66% accuracy, which is 10 percent lower in terms of absolute accuracy. This is probably rooted in the way the embedding vectors are generated. In the implementation, only the types of tokens are used in the word2vec algorithm. This may be enough for this model's original purpose, which is algorithm classification since same algorithms may have more structural similarities (e.g. same loop structure in bubble sort), however, as pointed out by the study of [7], the software commit bugs represented more of a "localness" characteristic, e.g. the spellings of certain tokens in the lines being more important. Filtering out this information may be crucial to this specific task.

## 5.3 The codeflaws dataset

The original dataset is presented in the form of 21 fault classes. Given that for some of these classes, the number of corresponding samples to the dataset is relevant small, we rearranged the dataset from 21 classes of correct/defect program pairs to just 2 classes of correct lines/defect lines.

### 5.3.1 TextCNN/Attention Model

For C files, we have trained a different embedding vector from these task-solving problems. It is worth noticing that in these programs the namings of the variables are usually more arbitrary than software packages that follows industrial standards. From the generated tsne [40] graphs indicating the distance between tokens, it is obvious that in the C embedding's case, the distance between common one-letter variables like i, j, k etc. are too close, which means the embedding contains less useful information. For input pairs, we used the lines that are faulty (changed) as the left hand side and the complete file in the right hand side.

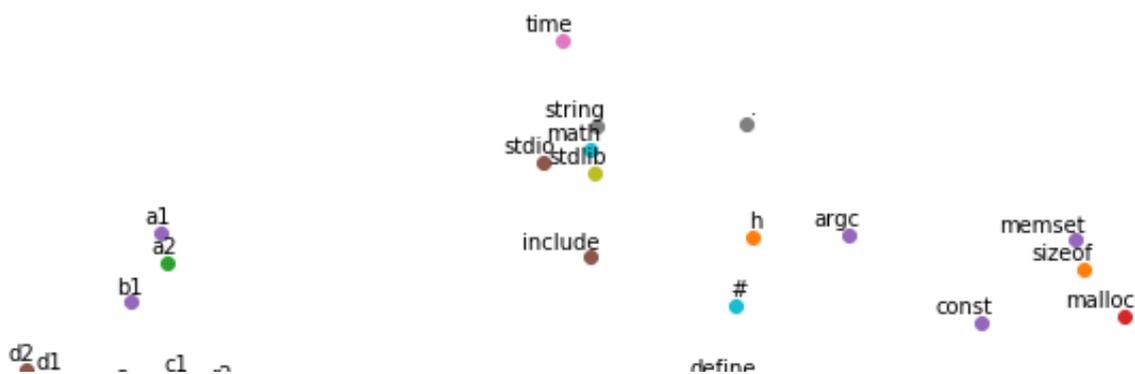

(a)

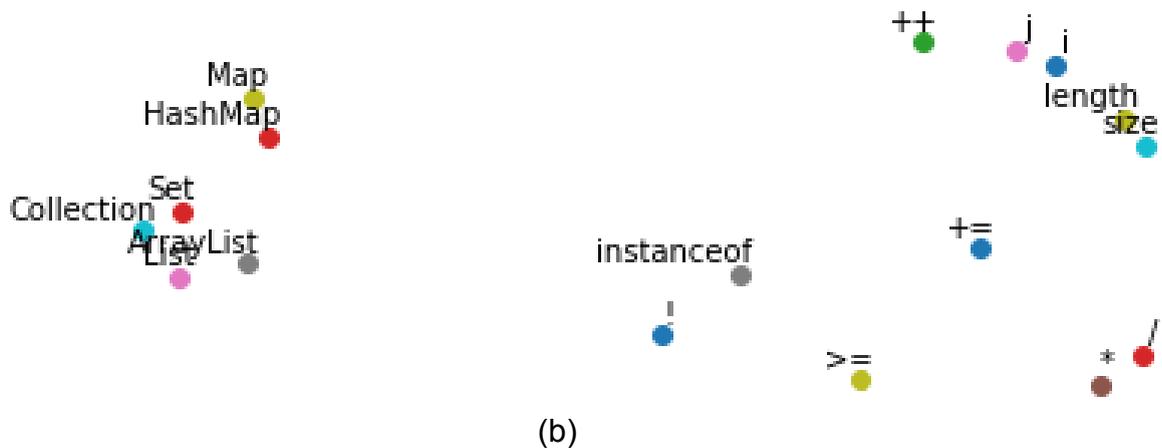

(b)

Fig 10. Part of the clusters in the generated TSNE graph from the embeddings from (a) C algorithm programs (b) Github Java corpus

### 5.3.2 TreeCNN Model

The TreeCNN model performed slightly better from others, with an accuracy of 52.74%.

## 5.4 The atcoder novice dataset

### 5.4.1 TextCNN/LSTM model

We are surprised to find out, that with a few lines of change in the code, there's a huge accuracy difference from the two models, although they are using the same embedding vectors. The TextCNN achieved 99.57% test accuracy, and on the contrary, the LSTM model achieved only 46.27%.

### 5.4.2 TreeCNN Model

The TreeCNN model achieved 90.69% test accuracy.

## 5.5 The GGNN Model

We have selected the existing implementation of GGNN from the particular branch that aims for algorithm classification[4] . We have re-generated two classes of sort algorithms on our machine and proved that the code is effective on two-way classifications. However, after feeding in our own datasets, the final accuracy always stays around 50%. We have concluded that the model is better at summarising the overall structure of the source code file rather than sequence/condition related properties.

---

[4] https://github.com/bdqnghi/ggnn_graph_classification

## 5.6 General language model

Our bert runtime environment is an AWS EC2 p2.xlarge instance with the Ubuntu Deep Learning AMI (this package provides ad-hoc cuda and deep learning library support). Given the GPU memory limitation, we have set the embedding dimension to 128 and number of layers to 4. Same as previous tasks, we trained the java model over the github java corpus. We have also reduced the number of epochs and the per-epoch size in training and tuning. Binary classification over sequence pairs has a similar structure as the "QNLI" task in the GLUE dataset. We therefore modify our dataset to fit the format of the QNLI task. Our data for reproduction is available at (https://github.com/hiroto-takatoshi/XLM). Over all , our tiny bert model has leveled or outscored previous models in this study.

| Dataset\Model | Bert Accuracy |
|---|---|
| Syntax Faults | 99.55 |
| Algorithm errors | 56.45 |
| General commits | 81.10 |

Table 4. Bert accuracy over defect tasks

# 6. Results Summary

## 6.1 Results

Table 5 summarises the experiment results. For existing models, it is obvious that in all these three levels of defects, there is at least one model that produces a promising outcome, but there is no clear pattern on which model suits what levels of the task.

| Dataset\Model | Sequential structures | | Tree/Graph Structures | | General LM |
|---|---|---|---|---|---|
| | TextCNN | LSTM | TreeCNN | GGNN | BERT(tiny) |
| Syntax Faults | 99.57 | 46.27 | 90.69 | Random output. Detailed discussion in previous sections | 99.55 |
| Algorithm errors | 49.01 | 50.60 | 52.74 | | 56.45 |
| General commits | 74.66 | 73.23 | 68.36 | | 81.10 |

Table 5. Test Accuracy (%) for each case

## 6.2 Findings

**RQ1. Are generally adapted language models in natural language tasks fitted for programming languages?**

With at least one model from other studies performing well in all three bug-finding tasks, we suggest that models for natural language can handle tasks for programming languages.

**RQ2. Are models designated for algorithm classification tasks fitted for defect classification task?**

In some tasks, the TreeCNN model performed well, but it is not the case with the graph model GGNN. We suggest that a tuning is needed with the GGNN model.

> **RQ3. For originally generated datasets, are they general enough to represent the overall feature of the task?**
>
> The general commit defect dataset, as suggested in previous chapters over accuracy difference when using different test datasets, is not general enough to represent the overall feature of software defect and needs refinement.

> **RQ4. Is there any common ground and similarity between the different levels of defects, so that a certain model is suited for all three tasks?**
>
> For models other than Bert, we have failed to find common patterns in the results, and there is not a model that achieves satisfying accuracy across three levels of defect.
>
> We are delighted that general language models like bert are possible to solve the various defect problems, but there is no accuracy patterns showing the complexity change between these levels.

> **RQ5. Is the training process in each task acceptable in terms of time and space efficiency?**
> Preprocessing the input data and training the neural models are all completed on a personal PC without extra hardware. Time cost of code maintenance and model training does not exceed a few hours.

## 6.3 Discussion

**Building a more general bug commit dataset.** Test results using test data from different repositories show that the training dataset does not contain enough features of bug commit. The original dataset from 10 repositories may not be enough. We suggest to build a corpus on a much larger scale in the future studies.
**Transform learning between datasets.** Transform learning is a tactic to tune an existing model to a new task that has similar structures. It is obvious that the three levels of defect in our study do share many properties in common, and training an existing model on another is a promising research aspect in the future.
**Bert for Java.** There has already been commercial attempts [41] to deploy the gpt2 model for automatic code completion tasks. Researchers have also been deploying general LMs to various fields [42] and measure its performance on a set of tasks.

Although our try on Bert is promising, we have not systematically evaluated our tiny bert model's performance over common SE tasks. We are suggesting to put further research focus on this direction.

# 7. Conclusion

This research studied the existing literature of programming language models - both linear and structural ones, as well as the literature of different level of software defects. These four models are then tested on three different datasets. In the experiment, input data are preprocessed and we managed to keep the original model unmodified. We have some surprising findings that some models have achieved an overwhelming accuracy, but we have failed to find common ground between the three levels of defects we defined. The results are analyzed, and we suggest to train a more general model, the "general language model" that can be fine-tuned into this specific task. Results show that our novel proposal of pre-training a general language model (Bert) outperforms existing models.

# 8. Acknowledgements

The research process is much assisted and instructed by Dr.Nicholas Matragkas. I would also like to thank the support from Cross Compass Inc. in Japan for providing peer review on the original proposal of deploying the Bert model.

(All appendixes should be marked)

# Appendixes

## Appendix 1 The atcoder dataset generation script

```python
#!/usr/bin/env python
from bs4 import BeautifulSoup
import requests
import re
import os
from tqdm import *

def parse_page(run_result="CE", page_num=1):
    pass

def parse_submission(suburl):
    base_sub_url = 'https://atcoder.jp'
    h = requests.get(base_sub_url + suburl).text
    soup = BeautifulSoup(h, 'html.parser')
    src = soup.find_all(class_='prettyprint')[0].get_text()
    return src

base_ce_url = 'https://atcoder.jp/contests/abs/submissions?f.Language=3016&f.Status=AC&f.Task=&f.User=&page='

CE_DIR = "C:/Users/admin/repo_for_lm/cf_java/NCE"

cnt = 0

for i in trange(1, 112):

    h = requests.get(base_ce_url + '1').text
    soup = BeautifulSoup(h, 'html.parser')

    for link in soup.find_all('a'):
        x = link.get('href')
        try:
            if "submissions/" in x and not 'me' in x:
```

```python
            src = parse_submission(x)
            cnt += 1
            with open(os.path.join(CE_DIR, str(cnt) + '.java'), 'w', encoding='utf8') as f:
                f.write(src)

    except TypeError as e:
        pass
```

## Appendix 2 C++ tokenization tool using Clang

```python
#!/usr/bin/env python
import clang.cindex

clang.cindex.Config.set_library_file('C:/Program Files/LLVM/bin/libclang.dll')

index = clang.cindex.Index.create()
tokens_list = list(index.parse('h.c').cursor.get_tokens())
for tok in tokens_list:
    print(tok.spelling, tok.location.line, tok.kind)
```

## Appendix 3 The sequential model's training script

```python
#!/usr/bin/env python
# coding: utf-8

# In[14]:

import tensorflow as tf
import numpy as np

# In[2]:

sess=tf.Session()
new_saver = tf.train.import_meta_graph('model.ckpt.meta')
new_saver.restore(sess, 'model.ckpt')
```

```python
graph = tf.get_default_graph()

# In[3]:

print(graph)

# In[3]:

[(n.op, n.name) for n in tf.get_default_graph().as_graph_def().node if "Variable" in n.op]

# In[7]:

vec = graph.get_tensor_by_name('Variable:0')

# In[8]:

norm = tf.sqrt(tf.reduce_sum(tf.square(vec), 1, keep_dims=True))

# In[9]:

normalized_embeddings = vec / norm

# In[11]:

final_embeddings = normalized_embeddings.eval(session=sess)

# In[14]:
```

```python
final_embeddings.shape
```

# In[1]:

```python
import tensorflow as tf
import numpy as np
import math
import collections
import random
import pickle
import glob,os
from tempfile import gettempdir
from sklearn.manifold import TSNE
import matplotlib.pyplot as plt
```

# In[4]:

```python
PATH_TO_STORE_THE_DICTIONARY="C:/Users/admin/chenw2k/dict_file"
```

# In[5]:

```python
with open(PATH_TO_STORE_THE_DICTIONARY , "rb") as f:
    [count,dictionary,reverse_dictionary,vocabulary_size] = pickle.load(f)
```

# In[9]:

```python
def plot_with_labels(low_dim_embs, labels, filename):
  assert low_dim_embs.shape[0] >= len(labels), 'More labels than embeddings'
  plt.figure(figsize=(18, 18))  # in inches
  for i, label in enumerate(labels):
    x, y = low_dim_embs[i, :]
    plt.scatter(x, y)
    plt.annotate(label,
```

```python
                    xy=(x, y),
                    xytext=(5, 2),
                    textcoords='offset points',
                    ha='right',
                    va='bottom')

    plt.savefig(filename)

tsne = TSNE(perplexity=30, n_components=2, init='pca', n_iter=5000, method='exact')
plot_only = 500
low_dim_embs = tsne.fit_transform(final_embeddings[:plot_only, :])
labels = [reverse_dictionary[i] for i in range(plot_only)]
plot_with_labels(low_dim_embs, labels, 'wtf1.png')
```

# In[21]:

```python
np.save('java_w2v', final_embeddings)
```

# In[1]:

```python
import numpy as np
from keras.preprocessing.text import Tokenizer
from keras.preprocessing.sequence import pad_sequences
from keras.utils import to_categorical
from keras.layers import *
from keras.models import Model, load_model
from keras.initializers import Constant, TruncatedNormal
from keras.callbacks import EarlyStopping, ModelCheckpoint, ReduceLROnPlateau
from keras.optimizers import Adam
from keras_self_attention import SeqSelfAttention

from sklearn.utils import shuffle
```

# In[2]:

```python
final_embeddings = np.load('emb_new.npy')

# In[3]:

import pickle

# In[4]:

f = open("bugs_new.pickle" , "rb")
[sent1, sent2, label] = pickle.load(f)

assert len(sent1) == len(sent2) and len(sent2) == len(label)

print(len(sent1))

# In[5]:

f2 = open("bugs_test.pickle" , "rb")
[s1, s2, lbl] = pickle.load(f2)

lbl = to_categorical(np.asarray(lbl))
s1 = pad_sequences(s1, maxlen=128)
s2 = pad_sequences(s2, maxlen=128)

s1, s2, lbl = shuffle(s1, s2, lbl)

# In[15]:

X_test = get_rnn_data(s1, s2)
Y_test = lbl

# In[6]:
```

```python
def get_rnn_data(a,b):
    x = {
        'sentence1': a,
        #
        'sentence2': b,
        }
    return x
```

# In[6]:

```python
ccc = 0
ddd = 0
for _ in sent1:
    for __ in _:
        if __ == 0:
            ccc += 1
        else: ddd += 1
print(ccc, ddd)
```

# In[7]:

```python
#print(label)

label = to_categorical(np.asarray(label))
sent1 = pad_sequences(sent1, maxlen=128)
sent2 = pad_sequences(sent2, maxlen=128)

sent1, sent2, label = shuffle(sent1, sent2, label)
```

# In[8]:

```python
X_train = get_rnn_data(sent1[:21000],sent2[:21000])
Y_train = label[:21000]
X_valid = get_rnn_data(sent1[21000:27000],sent2[21000:27000])
Y_valid = label[21000:27000]
```

```python
    print(label)

# In[9]:

X_test = get_rnn_data(sent1[27000:], sent2[27000:])
Y_test = label[27000:]

# In[ ]:

sent1, sent2, label = shuffle()

# In[10]:

X_train

# In[27]:

MAX_SEQUENCE_LENGTH = 128
EMBEDDING_DIM = 64
VOCABULARY_SIZE = 100000

embedding_layer = Embedding(VOCABULARY_SIZE,
            EMBEDDING_DIM,
            embeddings_initializer=Constant(final_embeddings),
            input_length=MAX_SEQUENCE_LENGTH,
            trainable=False)

def model_cnn(x):
    filter_sizes = [1,2,3,5]
    num_filters = 128

    x = Reshape((128, 64, 1))(x)

    maxpool_pool = []
```

```python
    for i in range(len(filter_sizes)):
        conv = Conv2D(num_filters, kernel_size=(filter_sizes[i], 64),
                      kernel_initializer='he_normal', activation='relu')(x)
        maxpool_pool.append(MaxPool2D(pool_size=(128 - filter_sizes[i] + 1, 1))(conv))

    z = Concatenate(axis=1)(maxpool_pool)
    z = Flatten()(z)
    z = Dropout(0.2)(z)

    return z

def toyCNN(x, col):
    x = Bidirectional(CuDNNLSTM(128, return_sequences=True))(x)
    u1 = SeqSelfAttention(attention_activation='sigmoid')(x)
    x = Bidirectional(CuDNNLSTM(128, return_sequences=True))(x)
    u2 = SeqSelfAttention(attention_activation='sigmoid')(x)
    x = Bidirectional(CuDNNLSTM(128, return_sequences=True))(x)
    u3 = SeqSelfAttention(attention_activation='sigmoid')(x)
    x = Bidirectional(CuDNNLSTM(128, return_sequences=True))(x)
    u4 = SeqSelfAttention(attention_activation='sigmoid')(x)

    x = concatenate([u1, u2, u3, u4])
    return x

inp1 = Input(shape=(128,), dtype='int32', name="sentence1")
inp2 = Input(shape=(128,), dtype='int32', name="sentence2")
emb1 = embedding_layer(inp1)
emb2 = embedding_layer(inp2)

#x = concatenate([toyCNN(emb1, "sent1"), toyCNN(emb2, "sent2")])
x = concatenate([model_cnn(emb1), model_cnn(emb2)])
preds = Dense(2, activation='sigmoid', name='densejoke')(x)
model = Model(inputs=[inp1,inp2], outputs=preds)

model.compile(loss='categorical_crossentropy',
        optimizer=Adam(lr=0.001),
        metrics=['acc'])

learning_rate_reduction = ReduceLROnPlateau(monitor='val_acc',
                            patience=1,
                            verbose=1,
```

```python
                         factor=0.5,
                         min_lr=0.00001)
file_path="checkpoint_SNLI_weights.hdf5"
checkpoint = ModelCheckpoint(file_path, monitor='val_acc', verbose=1,
save_best_only=True, mode='max', save_weights_only=True)
early = EarlyStopping(monitor="val_acc", mode="max", patience=3)

model_callbacks = [checkpoint, early, learning_rate_reduction]
```

# In[11]:

```python
model.load_weights(file_path)
```

# In[28]:

```python
model.fit(X_train, Y_train,
    batch_size=64,
    epochs=30,
    verbose=1,
    validation_data=(X_valid, Y_valid),
    callbacks = model_callbacks
    )
```

# In[29]:

```python
model.evaluate(X_test, Y_test, batch_size=64)
```

# In[14]:

```python
model.metrics_names
```

# In[33]:

```python
Y_nasha = model.predict(X_test, batch_size=64)
```

# In[34]:

```python
Y_nasha = np.argmax(Y_nasha,axis=1)
```

# In[35]:

```python
Y_test_nasha = np.argmax(Y_test, axis=1)
```

# In[36]:

```python
def count0(l):
    r = 0
    for _ in l:
        if _ == 0: r += 1
    return r * 1.0 / len(l)
```

# In[37]:

```python
ssb = []

for i, (x, y) in enumerate(zip(Y_nasha, Y_test_nasha)):
    if x != y:
        ssb.append(count0(X_test['sentence1'][i]))
        ssb.append(count0(X_test['sentence2'][i]))
```

# In[38]:

```python
import matplotlib.pyplot as plt
```

```
plt.hist(ssb,  bins=[0.0, 0.1, 0.2, 0.3, 0.4, 0.5, 0.6, 0.7, 0.8, 0.9, 1.0])  # arguments are passed to np.histogram
plt.title("Histogram with 'auto' bins")
plt.show()
```

# In[43]:

```
kkb = []

for i, (x, y) in enumerate(zip(sent1, sent2)):
    kkb.append(count0(x))
    kkb.append(count0(y))

plt.hist(kkb,  bins=[0.0, 0.1, 0.2, 0.3, 0.4, 0.5, 0.6, 0.7, 0.8, 0.9, 1.0])  # arguments are passed to np.histogram
plt.title("Histogram with 'auto' bins")
plt.show()
```

# In[ ]:

## Appendix 4 Modified Bi-TBCNN Code

```
# This file is just another version to test with 2 different AST tree on each side of the Bi-TBCNN
import os
import logging
import pickle
import tensorflow as tf
import numpy as np
import network as network
import sampling as sampling
from parameters import LEARN_RATE, EPOCHS, CHECKPOINT_EVERY, BATCH_SIZE, DROP_OUT, TEST_BATCH_SIZE
from sklearn.metrics import classification_report, confusion_matrix, accuracy_score
```

```python
import random
import sys
import json
import argparse

# os.environ['CUDA_VISIBLE_DEVICES'] = '0,1,2,3,4,5,6,7'
#os.environ['CUDA_VISIBLE_DEVICES'] = "-1"
os.environ['TF_CPP_MIN_LOG_LEVEL'] = '3'

LABELS_bak = ['OAAN', 'OLLN', 'SISF', 'SISA', 'OMOP', 'SRIF', 'HBRN', 'OAID', 'DCCR', 'SMOV',\
    'OITC', 'OAIS', 'OIRO', 'DRWV', 'SDIF', 'SDIB', 'OEDE', 'SIRT', 'DMAA', 'ORRN', 'SIIF',\
    'OICD', 'STYP', 'DCCA', 'HCOM', 'DRAC', 'SMVB', 'SDFN', 'HDMS', 'HDIM', 'OFFN', \
    'DRVA', 'OILN', 'HIMS', 'OFPF', 'SDLA', 'OFPO', 'HEXP', 'HOTH']

LABELS = ['bug', 'norm']

# device = "/cpu:0"
# device = "/device:GPU:0"
def get_one_hot_similarity_label(left_labels, right_labels): # func modified for multi cases
    sim_labels = []
    sim_labels_num = []
    for i in range(0,len(left_labels)):

        label_vec = np.zeros(len(LABELS))
        label_vec[LABELS.index(left_labels[i])] = 1.0
        #print(label_vec)
        #exit(0)
        sim_labels.append(label_vec)

    return sim_labels, []

def get_trees_from_pairs(label_1_pairs,labeL_0_pairs):
    all_pairs = label_1_pairs + labeL_0_pairs
    random.shuffle(all_pairs)
    left_trees = []
```

```python
        right_trees = []
        for pair in all_pairs:
            left_trees.append(pair[0])
            right_trees.append(pair[1])
        return left_trees, right_trees

def trees_from_pairs(all_pairs):
    random.shuffle(all_pairs)
    left_trees = []
    right_trees = []
    for pair in all_pairs:
        left_trees.append(pair[0])
        right_trees.append(pair[1])
    return left_trees, right_trees

def generate_random_batch(iterable,size):
    l = len(iterable)
    for ndx in range(0, l, n):
        yield iterable[ndx:min(ndx + n, l)]

def train_model(logdir, inputs, left_embedfile, right_embedfile, epochs=EPOCHS, with_drop_out=1,device="-1"):
    os.environ['CUDA_VISIBLE_DEVICES'] = device

    print("Using device : " + device)
    print("Batch size : " + str(BATCH_SIZE))
    if int(with_drop_out) == 1:
        print("Training with drop out rate : " + str(DROP_OUT))
    n_classess = len(LABELS)
    left_algo_labels = LABELS
    right_algo_labels = LABELS

    # left_algo_labels = ["bfs","bubblesort","knapsack","linkedlist","mergesort","quicksort"]
    # right_algo_labels = ["bfs","bubblesort","knapsack","linkedlist","mergesort","quicksort"]
    # with open(left_inputs, 'rb') as fh:
    #     left_trees, _, left_algo_labels = pickle.load(fh)

    # with open(right_inputs, 'rb') as fh:
```

```python
#     right_trees, _, right_algo_labels = pickle.load(fh)
print("Loading training data....")
# print "Using device : " + device
with open(inputs, "rb") as fh:
    all_1_pairs, all_0_pairs = pickle.load(fh)

# proportion for trian/valid dataset

print(len(all_1_pairs), len(all_0_pairs))
#exit(0)

valid_1_pairs = all_1_pairs[2800:3500]
#valid_0_pairs = all_0_pairs[10000:13000]

all_1_pairs = all_1_pairs[:2800]
#all_0_pairs = all_0_pairs[:10000]

left_trees, right_trees = trees_from_pairs(valid_1_pairs) #modified
#print(len(valid_1_pairs), len(left_trees))
#exit(0)

# print "Shuffling training data"
# random.shuffle(all_1_pairs)
# random.shuffle(all_0_pairs)

print("Loading embdding vectors....")
with open(left_embedfile, 'rb') as fh:
    left_embeddings, left_embed_lookup = pickle.load(fh)

with open(right_embedfile, 'rb') as fh:
    right_embeddings, right_embed_lookup = pickle.load(fh)

num_feats = len(left_embeddings[0])

# build the inputs and outputs of the network
left_nodes_node, left_children_node, left_pooling_node = network.init_net_for_siamese(
    num_feats
    )
```

```python
    right_nodes_node, right_children_node, right_pooling_node = network.init_net_for_siamese(
        num_feats
    )
    # with tf.device(device):
    merge_node = tf.concat([left_pooling_node, right_pooling_node], -1)

    hidden_node = network.hidden_layer(merge_node, 200, 200)
    if int(with_drop_out) == 1:
        hidden_node = tf.layers.dropout(hidden_node, rate=DROP_OUT, training=True)

    hidden_node = network.hidden_layer(hidden_node, 200, 200)

    if int(with_drop_out) == 1:
        hidden_node = tf.layers.dropout(hidden_node, rate=DROP_OUT, training=True)

    hidden_node = network.hidden_layer(hidden_node, 200, n_classess)

    if int(with_drop_out) == 1:
        hidden_node = tf.layers.dropout(hidden_node, rate=DROP_OUT, training=True)

    out_node = network.out_layer(hidden_node)

    labels_node, loss_node = network.loss_layer(hidden_node, n_classess)

    learning_rate = tf.placeholder(tf.float32, shape=[])
    optimizer = tf.train.AdamOptimizer(learning_rate=learning_rate)
    train_step = optimizer.minimize(loss_node)

    # tf.summary.scalar('loss', loss_node)

    # correct_prediction = tf.equal(tf.argmax(out_node,1), tf.argmax(labels_node,1))
    # accuracy = tf.reduce_mean(tf.cast(correct_prediction, tf.float32))

    ### init the graph
    # config = tf.ConfigProto(allow_soft_placement=True)
    # config.gpu_options.allocator_type = 'BFC'
    # config.gpu_options.per_process_gpu_memory_fraction = 0.9
```

```python
# config = tf.ConfigProto()
# config.gpu_options.allow_growth = True

config = tf.ConfigProto()
config.gpu_options.allocator_type ='BFC'
# config.gpu_options.allow_growth = True
config.gpu_options.per_process_gpu_memory_fraction = 0.98

sess = tf.Session(config = config)

# sess = tf.Session()
sess.run(tf.global_variables_initializer())

with tf.name_scope('saver'):
    saver = tf.train.Saver()
    summaries = tf.summary.merge_all()
    writer = tf.summary.FileWriter(logdir, sess.graph)
    ckpt = tf.train.get_checkpoint_state(logdir)
    if ckpt and ckpt.model_checkpoint_path:
        print("Continue training with old model")
        saver.restore(sess, ckpt.model_checkpoint_path)
    # else:
    #     raise 'Checkpoint not found.'

checkfile = os.path.join(logdir, 'cnn_tree.ckpt')
steps = 0

using_vector_lookup_left = False
if os.path.isfile("/input/config.json"):
    file_handler = open(config_file, 'r')
    contents = json.load(file_handler)
    using_vector_lookup_left = contents['using_vector_lookup_left'] == "false"

print("Begin training....")

# with tf.device(device):

# dumb implement the reduce on plateau func
previous_loss = 0.0
rounds_no_improvement = 0
```

```python
    lr = LEARN_RATE

    loss_list = []

    def run_validation():
        correct_labels = []
        predictions = []
        print('Computing validation accuracy...')
        print(len(left_trees))
        sum_ones = 0
        correct_ones = 0
        for left_gen_batch, right_gen_batch in sampling.batch_random_samples_2_sides(left_trees, left_algo_labels, right_trees, right_algo_labels, left_embeddings, left_embed_lookup, right_embeddings, right_embed_lookup, using_vector_lookup_left, False, TEST_BATCH_SIZE):
            #print("i am here")
            try:
                left_nodes, left_children, left_labels_one_hot, left_labels = left_gen_batch

                right_nodes, right_children, right_labels_one_hot, right_labels = right_gen_batch
                sim_labels, _ = get_one_hot_similarity_label(left_labels,right_labels)
                #print(("sim labels : " + str(sim_labels)))
                output = sess.run([out_node],
                    feed_dict={
                        left_nodes_node: left_nodes,
                        left_children_node: left_children,
                        right_nodes_node: right_nodes,
                        right_children_node: right_children,
                        labels_node: sim_labels
                    }
                )
                correct = np.argmax(sim_labels[0])
                predicted = np.argmax(output[0])
                check = (correct == predicted) and True or False
                if check: correct_ones += 1
                sum_ones += 1
                #print(('Out:', output, "Status:", check, "acc: ", correct_ones / sum_ones))
                correct_labels.append(np.argmax(sim_labels[0]))
                predictions.append(np.argmax(output[0]))
            except Exception as err_msg:
                #print(err_msg)
```

```python
            pass

        target_names = ["0","1"]
        ret_acc = accuracy_score(correct_labels, predictions)
        print('Validation Accuracy:', ret_acc)
        return ret_acc

    for epoch in range(1, epochs+1):

        if epoch % 10 == 0:
            loss_list = []

        print("epoch ", epoch)

        #optimizer = tf.train.AdamOptimizer(lr)
        #train_step = optimizer.minimize(loss_node)

        SAMPLE_PAIR_SIZE = 100
        sample_1_pairs = random.sample(all_1_pairs,SAMPLE_PAIR_SIZE)
        #sample_0_pairs = random.sample(all_0_pairs,SAMPLE_PAIR_SIZE)
        shuffle_left_trees, shuffle_right_trees = trees_from_pairs(sample_1_pairs)
        print("Left left:",len(shuffle_left_trees),"Len right:",len(shuffle_right_trees))
        for left_gen_batch, right_gen_batch in sampling.batch_random_samples_2_sides(shuffle_left_trees, left_algo_labels, shuffle_right_trees, right_algo_labels, left_embeddings, left_embed_lookup, right_embeddings, right_embed_lookup, using_vector_lookup_left, False, BATCH_SIZE):
            #print("-----------------------------------------------------")
            left_nodes, left_children, left_labels_one_hot, left_labels = left_gen_batch

            right_nodes, right_children, right_labels_one_hot, right_labels = right_gen_batch

            sim_labels, sim_labels_num = get_one_hot_similarity_label(left_labels,right_labels)
            #print(sim_labels)

            _, err, out, merge, labs, left_pooling = sess.run(
                [train_step, loss_node, out_node, merge_node, labels_node, left_pooling_node],
                feed_dict={
```

```
                    learning_rate: lr,
                    left_nodes_node: left_nodes,
                    left_children_node: left_children,
                    right_nodes_node: right_nodes,
                    right_children_node: right_children,
                    labels_node: sim_labels
                }
            )

            # print "hidden : " + str(loss)
            #print('Epoch:', epoch,'Steps:', steps,'Loss:', err)
            #loss_list.append(err)

            if steps % CHECKPOINT_EVERY == 0:
                # save state so we can resume later
                saver.save(sess, os.path.join(checkfile), steps)
                print('Checkpoint saved.')

            steps+=1

        if epoch > 0 and epoch % 30 == 0:
            valid_acc = run_validation()
            if valid_acc <= previous_loss:
                rounds_no_improvement += 1
                if rounds_no_improvement > 3:
                    print("no further improvements, ending...")
                    break
                lr = lr * 0.5
                print("learning rate reduced to ", lr)
            else:
                rounds_no_improvement = 0
                previous_loss = valid_acc

        steps = 0

def main():

    # example params :
        # argv[1] = ./bi-tbcnn/bi-tbcnn/logs/1
        # argv[2] = ./sample_pickle_data/all_training_pairs.pkl
        # argv[3] = ./sample_pickle_data/python_pretrained_vectors.pkl
```

```
        # argv[4] = ./sample_pickle_data/fast_pretrained_vectors.pkl
        # argv[5] = 1
    train_model(sys.argv[1],sys.argv[2],sys.argv[3], sys.argv[4], 1000, 0, "0")

if __name__ == "__main__":
    main()
```